\documentstyle[epsf]{article}

 {\par\noindent{\underline{Proof} \quad}}{\hfill$\Box$\bigskip}
 {\par\noindent{\underline{Proof} of the theorem\quad}}{\hfill$\Box$\bigskip}
 {\par\smallskip\noindent{\underline{{\it Remark}} \quad}}{\par\smallskip}
 {\par\smallskip\noindent{\underline{{\it Fact}} \quad}}{\par\smallskip}
 {\par\smallskip\noindent{\underline{{\it Example}} \quad}}{\par\smallskip}
 {\par\smallskip\noindent{{\it Assumotion} \quad}}{\par\smallskip}
 {\par\smallskip\noindent{{\it Condition} \quad}}{\par\smallskip}

\begin{document}
\title{NMR C-NOT gate through Aharanov-Anandan$'$s phase shift   }
\author{Wang Xiangbin\thanks{email: wang@qci.jst.go.jp}, Matsumoto 
Keiji\thanks{email: keiji@qci.jst.go.jp} \\
        Imai Quantum Computation and Information project, ERATO, Japan Sci. and Tech. Corp.\\
Daini Hongo White Bldg. 201, 5-28-3, Hongo, Bunkyo, Tokyo 113-0033, Japan}

\maketitle
\begin{abstract} 
Recently, it is proposed to do quantum computation through the Berry$'$s 
phase( adiabatic cyclic geometric phase) shift with NMR
( Jones {\it et al} {\it Nature} {\bf 403}, 869(2000)). 
This geometric quantum gate is hopefully to be fault 
tolerant to certain types of errors because
of the geometric property of the Berry phase.  
Here we give a  scheme to realize the NMR C-NOT gate through Aharonov-Anandan$'$s
 phase( non-adiabatic cyclic phase) shift on the dynamic phase free evolution loop.
 In our scheme, the gate is run non-adiabatically, thus it
is less affected by the decoherence. And, in the scheme we have chosen the
the zero dynamic phase time evolution loop in obtaining the gepmetric phase shift, 
we need not take any extra operation to cancel the dynamic phase.   
\end{abstract}
\newpage
It has been shown that, a quantum computer, if available, can perform certain tasks much more
efficiently than a classical Turing machine\cite{loyd, bennett, di}. The realization of the basic
constitute of quantum computer, namely fault-toleratant quantum logic gate, is a certral issue.
It has been shown that, with the single qubit rotation
operation and the C-NOT gate, 
one can in principle realize arbitrary quantum computation\cite{bar}.

Geometric 
phase\cite{panch, berry, aha} plays
an important role in quantum interferometry and many other disciplines. 
For the two level system, Geometric phase is equal to half of the solid angle subtended
by the area in the Bloch sphere enclosed by the closed evolution loop of the eigenstate.
Recently\cite{ekert}, it was proposed to make a fault tolerate C-NOT gate using Berry$'s$ phase\cite{berry}, i.e.,
adiabatic and cyclic geometric phase.  Similar idea was then developed to asymmetric SQUID system\cite{vedral}.
However, the proposal\cite{ekert} relies on the adiabatic operations.     
As has been shown in \cite{ekert}, due to its geometric property, 
geometric phase is fault tolerant to certain types of
operational errors. So the proposals of quantum computation through 
conditional geometric phase shift are potentially important in the future
implementation of a fault tolerate quantum computer. 
However, one should overcome two drawbacks 
in the previous suggestions\cite{ekert, vedral} 
in doing the geometric quantum computation. 
One is the adiabatic condition which
makes such gate not practical.
The decoherence effects may distort the laboratory observations seriously. 
The experimental results with systematic errors were obtained on NMR\cite{ekert}. 
It is also reported that the distortion is increased seriously with the faster running speed. 
 The other drawback is the extra operation to eliminate the dynamic phase. 
This extra operation  weakens the fault tolerate property\cite{alexandre}.
In this letter, we give a 
simple scheme to realize the geometric 
C-NOT gate non-adiabatically on the $0$ dynamic phase evolution curve
with  NMR. 
By this scheme, the above mentioned two drawbacks of previous suggestions are removed
and the realization can be done faster and more easily.  
We believe our scheme has led the idea of geometric C-NOT gate much more practical than before.

Geometric phase also exists in non-adiabatic process.     
It was  shown by Aharonov and Anandan\cite{aha} that the
geometric phase is only dependent on the area enclosed by the loop of the 
state on the Bloch sphere. In non-adiabatic case, the path of the state evolution in general is 
different from the path of the parameters in the Hamiltonian. 
The external field 
need not always follow the evolution path of the state like that in adiabatic case. 
So it is possible to let the external field
perpendicular to the evolution path instantaneously so that there is no dynamic phase 
involved in the whole process.
             
The detection of AA phase was sucessfully done many years ago\cite{suter}. In particular,
one may select the geodesic evolution loop so that no dynamic phase is involved in the interference
result.
It should be interesting to develop a non-adiabatic and dynamic phase free 
scheme to realize the two qubits C-NOT
gate through geometric phase shift with NMR\cite{ekert}. 
The scheme\cite{suter} for non-adiabatic detection of geometric phase to single qubit can be easily developed for 
two qubits system to make a 
C-NOT gate with NMR. 

Consider the interacting nucleus spin pair(spin $a$ and spin $b$) in the NMR 
quantum computation\cite{cory, gers, jones, jones1}. 
For simplicity we call them as qubit $a$ and qubit $b$, respectively. 
We will use subscripts $a$ and $b$
to indicate the corresponding qubits.
If there is no horizontal field the Hamiltonian for the two bits system is
$H_i=\frac{1}{2}(\omega_{a} \sigma_{za}+\omega_b\sigma_{zb}+ J \sigma_{za}\cdot\sigma_{zb}) $,
where $\omega_{a,b}$ is the resonance frequency for spin $a,b$ respectively in a very high $+z$ direction
static magnetic field(e.g. $\omega_a$ can be $500$MHz\cite{ekert}) ,  $J$ is the interacting constant between
nucleus and $\sigma_{za}=\sigma_{zb}=\sigma_z=\left(\begin{array}{cc}
1&0\\0&-1\end{array}\right)$.  
The Hamiltonian for spin $a$ in the rotational framework  of rotating speed $\omega_a'=\omega_a-J$ is   
\begin{eqnarray}
H_a=R'H'R'^{-1}+i(\partial R'/\partial t)R'^{-1}=\frac{1}{2}(\omega_a-\omega_a'\pm J)\sigma_z
\end{eqnarray}
and $R'=e^{i\omega_a'\sigma_zt/2}$. Note here $H_a$ is dependent on  state of spin $b$ 
through $\pm J$. 
Explicitly, it is $\frac{1}{2}\cdot (2J) \sigma_z$ if state of qubit $b$ is 
$|\psi_b>=|\uparrow>$, and {\it it is $0$} 
if state of qubit $b$ is $|\psi_b>=|\downarrow>$.

Suppose we have chosen $\omega_a$  obviously different from $\omega_b$.
This means while we take operations on qubit $a$, state of qubit $b$ is (almost) not affected.  

\begin{figure}
\begin{center}
\epsffile{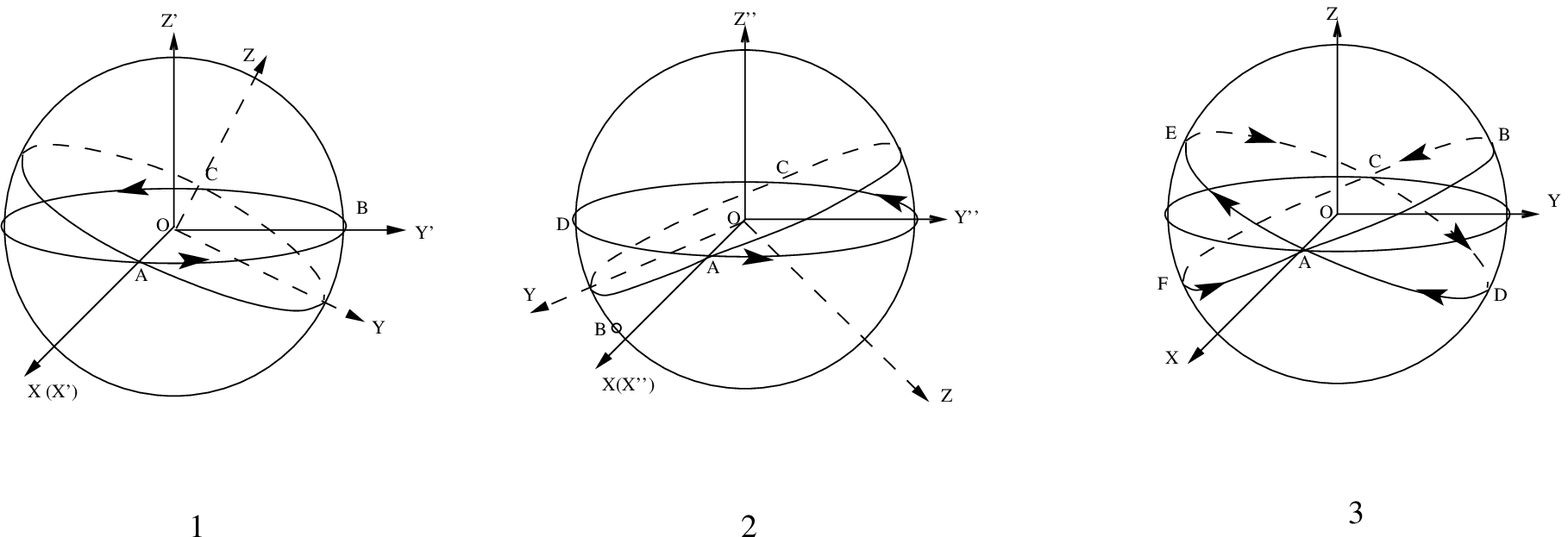}
\end{center}
\caption{ {\bf Non-adiabatic conditional geometric phase shift acquired through $0$ dynamic phase
evolution path}. These are pictures for the time evolution on Bloch sphere of qubit $a$ 
 in the case that qubit $b$ is $|\uparrow>$.
   Picture $1$ shows that after the Bloch sphere is rotated around
$x-$axis for $-\theta$ angle, the interaction Hamiltonian will rotate 
the 
Bloch sphere around $z'-$axis. At the time it completes $ \pi$  rotation,
i.e. $\tau=\pi/(2J)$
we rotate the bloch sphere around $x-$axis again for an angle of $-(\pi-2\theta)$,
 then we 
get  picture ${ 2}$. In picture ${2}$ the sphere is rotated around
$z''-$axis by the interaction Hamiltonian. Note that point $B$ in picture
$b$ has changed its position now. The geodesic cure CBA is not drawn
in picture $2$. After  time $\tau$ we
rotate the qubit $a$ around $x-$axis for an angle
of $\pi-\theta$.} Picture $3$  shows the whole evolution path on the Bloch sphere. 
point $A$ evolves
along closed curve ABCDA, a geometric phase $\gamma=-2\theta$ is acquired. 
Point $C$ evolves along the 
loop   CFAEC, a geometric phase $-\gamma=2\theta$ is acquired.
These are pictures for the time evolution on Bloch sphere of qubit $a$ 
$only$ in the
case that
qubit $b$ is $|\uparrow>$. If qubit $b$ is $|\downarrow>$, after the operation, qubit $a$
comes back to its initial state exactly.

\label{cnot}    \end{figure}

We first 
rotate qubit $a$ around  $x-$axis for angle $-\theta$ see fig.( \ref{cnot}). 
This operation  is denoted by $(-\theta)^x$. 
Note that $|\omega_b-\omega_a'|$ is much larger than  $J$,   so
the state of qubit $b$ is (almost) not affected by any operation on qubit $a$ in the whole 
process. 
The interaction Hamiltonian will create an evolution path on the geodesic circle 
ABC( see fig. \ref{cnot}).
After time 
$\tau=\pi/(2J)$,  we rotate qubit $a$ around $x-$axis for another angle $-(\pi-2\theta)$. 
Again wait for a  time $\tau$.
Then rotate qubit $a$ around $x-$axis for angle $\pi-\theta$ to let the Bloch sphere back 
to the original
one. 
In short, the above scheme can be expressed as
\begin{eqnarray}
\hat S=(-\theta)^x\longrightarrow \tau\longrightarrow [-(\pi-2\theta)]^{x}\longrightarrow \tau
\longrightarrow (\pi-\theta)^{x}
\end{eqnarray}
In the scheme we have used the rotation operation to qubit $a$, around $x-$axis. This can be done
by rf pulse.

After the above operation, if qubit $b$ is on state $|\uparrow>$,
an evolution path of ABCDA or CFAEC on the Bloch sphere is produced for qubit $a$; 
if qubit $b$ is on state 
$|\downarrow>$, nothing happens to qubit $a$. 
This is equivalent to say the time evolution
operator has the property $U(2\tau)|\pm>=e^{\pm \gamma(\theta) }|\pm>$ if qubit $b$ is up and
$U(2\tau)=1$ if qubit $b$ is down. Here states $|\pm>$ correspond to point A and C respectively
in Bloch sphere. $\gamma(\theta)$ is the geometric phase acquired 
for initial state 
$|+>$(point $A$ in
fig.\ref{cnot}). It is the half solid angle subtended by the area
ABCDA and $\gamma(\theta)=-2\theta$. 
In the basis of 
$|\uparrow>$ and $|\downarrow>$(eigenstste of $\sigma_z$), 
if qubit $b$ is up we have the following time evolution formula for 
qubit $a$:
\begin{eqnarray}
U(2\tau)\left(\begin{array}{c}|\uparrow>\\|\downarrow>\end{array}\right)
=\left(\begin{array}{cc}\cos \gamma(\theta)& i\sin\gamma(\theta)\\
 \\ i\sin\gamma(\theta) & \cos\gamma(\theta)
\end{array}\right)\left(\begin{array}{c}|\uparrow>\\|\downarrow>\end{array}\right).
\end{eqnarray}
 Note if qubit $b$ is down, there is no change 
to qubit $a$, in any basis.
We see $|\gamma(\theta)|=\pi/2$ makes a C-NOT gate here( see fig\ref{cnot}). This corresponds to 
$\theta=\pi/4$. Note here the state of qubit $a$ itself ingeneral 
does not undergo a cyclic evolution. However, the evolution path of qubit $a$ is completely determined
by $\gamma$, which is the AA phase of state $|+>$.  

In summary we have a scheme that can be used to make a C-NOT gate through pure geometric phase, 
which is fault tolerate 
to certain types of errors\cite{ekert}. This $\gamma$ is fault tolerant to those errors
which do not change the area enclosed by evolution loop of states $|\pm>$.
Here we need not take any extra action to remove the dynamic phase\cite{ekert, vedral}.
And also in our scheme
the operation
is done non-adiabatically so the running speed of our geometric gate is at the same level of 
the normal gate. 
We believe our scheme has  led the idea of geometric C-NOT gate\cite{ekert} to be 
much more closer to practical use. 

{\bf Acknowledgement:} We thank Prof Imai for support. WXB thanks Prof A. Ekert(Oxford) for fruitful discussions.

\newpage
\end{document}